\documentclass[twocolumn,preprintnumbers,amsmath,amssymb,floatfix,superscriptaddress,nofootinbib]{revtex4}
\pdfoutput=1
\usepackage{amsmath,hyperref,amssymb}
\usepackage{graphicx}
\usepackage{dcolumn}
\usepackage{bm}
\usepackage{verbatim}
\usepackage{tabularx}
\usepackage{slashed}
\usepackage{cancel}
\usepackage[boxsize=2em,aligntableaux=center]{ytableau}

\def\be{\begin{equation}}
\def\ee{\end{equation}}
\def\bea{\begin{eqnarray}}
\def\eea{\end{eqnarray}}

\newcommand{\yss}{\ytableausetup{boxsize=0.5em}}
\newcommand{\ysn}{\ytableausetup{boxsize=2em}}
\newcommand{\symm}{\yss\ydiagram{2}\ysn}
\newcommand{\asymm}{\yss\ydiagram{1,1}\ysn}
\newcommand{\fund}{\yss\ydiagram{1}\ysn}

\newcommand{\mc}{\mathcal}
\renewcommand{\t}{\tilde}

\begin{document}

\preprint{SU-ITP-13/17, SLAC-PUB-15726}

\title{S-duality of nonsupersymmetric gauge theories}


\author{Anson Hook}
\affiliation{School of Natural Sciences,
 Institute for Advanced Study\\
Princeton, NJ 08540, USA}
\author{Gonzalo Torroba}
\affiliation{Stanford Institute for Theoretical Physics and SLAC\\
 Stanford, CA 94305, USA}


\begin{abstract}
We propose a method for constructing pairs of nonsupersymmetric gauge theories related by S-duality. Starting from a known S-duality of supersymmetric theories realized on the worldvolume of D3 branes in type IIB string theory, a new duality is obtained by replacing the D3-branes with antibranes. Large classes of dual pairs of nonsupersymmetric theories can be obtained in this way, with different interactions and matter content (chiral and vector-like). The approach is illustrated on gauge theories realized on three-branes and fractional branes probing orbifold singularities. 
The duality sheds light on the dynamics of gauge theories and their possible infrared phases by providing concrete magnetic dual descriptions of strongly coupled theories.
Some of the models share various properties with QCD, including confinement and chiral symmetry breaking. More generally, these theories feature fermions in multiple two-index representations and could realize intriguing phases such as a free magnetic phase with chiral symmetry breaking or mixed phases where an interacting fixed point coexists with a confining phase. 
\end{abstract}

\maketitle


\section{Introduction}\label{sec:intro}

One of the central problems of quantum field theory is to understand the dynamics of gauge theories, especially the mechanism responsible for confinement. An early and intriguing proposal was that confinement could be formulated as the dual Meissner effect for magnetic monopoles~\cite{Mandelstam:1974pi}. The idea that gauge theories could admit dual descriptions motivated some of the fundamental discoveries
in theories with supersymmetry, including the Seiberg-Witten solution of $\mc N=2$ super Yang-Mills (SYM)~\cite{Seiberg:1994rs} and Seiberg duality in $\mc N=1$ super QCD~\cite{Seiberg:1994pq}. The main virtue of supersymmetry is that it can provide analytic control at strong coupling. One could then hope that some of the phenomena discovered with the help of supersymmetry are valid more generally.
Unfortunately, it has proven quite difficult to extend the supersymmetric results to nonsupersymmetric theories. 

A possible approach along these lines is to identify nonsupersymmetric theories that inherit their dynamical properties from supersymmetric `parent' theories. An example of this, which shares some similarities with our proposal, is the orbifold projection introduced originally by~\cite{Kachru:1998ys}. There, discrete $\mathbb Z_k$ projections of $\mc N=4$ SYM with varying degrees of supersymmetry inherit at large $N$ their conformal dynamics and holographic duality from the maximally supersymmetric theory. This technique can be extended to obtain nonsupersymmetric dualities by projecting from known supersymmetric dual pairs~\cite{Schmaltz:1998bg}. By now, many orbifold and orientifold equivalences have been studied, as reviewed in~\cite{Armoni:2004uu} and references therein. The equivalence is valid only at large $N$ and requires that the discrete symmetries  are not dynamically broken in the IR~\cite{Kovtun:2004bz}. A related development is the large $N$ volume independence of~\cite{Kovtun:2007py}, which allows one to understand the dynamics of nonsupersymmetric theories with unbroken center symmetry, such as QCD with adjoints.

In order to improve our understanding of nonsupersymmetric gauge theories,
it is important to find dualities that hold exactly and whose validity is independent of the presence of supersymmetry, discrete symmetries or large $N$. In fact, string theory provides such a candidate: the $SL(2,\mathbb Z)$ symmetry of the type IIB theory is believed to be exact~\cite{Polchinski:1998rr}, even in nonsupersymmetric solutions. The generator $S$ of this discrete group gives rise to strong/weak dualities in gauge theories that can be realized on the worldvolume of D3 branes. For D3 branes in flat space, this is the Montonen-Olive duality for $\mc N=4$ SYM~\cite{Witten:1998xy}, and in three dimensional brane models $S$ can be identified with mirror symmetry for 3d supersymmetric theories~\cite{Intriligator:1996ex,Hanany:1996ie}. The D-brane derivation of these dualities is based on the self-duality of D3-branes, a property that will also be crucial to our construction.

In this work, we propose a mechanism for obtaining dualities in nonsupersymmetric gauge theories based on replacing D3 branes by \textit{anti}--D3 branes ($ \overline{D3}$) and applying the $S$-duality of type IIB string theory. We start from a pair of supersymmetric theories obtained from $S$-duality on a system of D3 branes (and other ingredients discussed below) and  replace $D3 \to \overline{D3}$. This breaks supersymmetry explicitly at the string scale, but since $\overline{D3}$ branes are also self-dual, the string theory $S$-duality will map the system to another known configuration with $\overline{D3}$ branes. In an appropriate low energy limit,
the degrees of freedom on the D-brane worldvolume decouple from the gravitational bulk, and the string duality is expected to reduce to a gauge theory duality.  In this approach, the nonsupersymmetric duality is inherited from the supersymmetric $S$-dual pair, which is useful for identifying and testing the duality.
(It may also be possible to consider theories which are not of this type, like nonsupersymmetric orbifolds). The duality is expected to be valid also at small $N$, away from the planar limit.

An immediate motivation for our investigation comes from~\cite{Uranga:1999ib,Sugimoto:2012rt}, who applied $S$-duality to the $O3-\overline{D3}$ system obtaining a nonsupersymmetric version of the Montonen-Olive duality of $\mc N=4$ gauge theories. The duality in this setup provides a realization of confinement as a dual Meissner effect in a nonsupersymmetric context~\cite{Sugimoto:2012rt}. 
Here we will study the consequences of $S$-duality for more general $\overline{D3}$ brane configurations, with explicit examples from branes near singularties based on the supersymmetric duality of~\cite{GarciaEtxebarria:2012qx,Garcia-Etxebarria:2013tba}.

The dualities constructed in this way have various interesting features. First, concrete examples will exhibit an
 $S$-duality that acts as a strong/weak duality between an ``electric'' theory that becomes strongly coupled in the IR, and a ``magnetic'' IR free description. Also, some of the models are quite similar to QCD; they confine and break chiral symmetry. Another interesting aspect of these theories is that they naturally contain fermions in multiple two-index representations, coming from the gauginos plus additional fermionic flavors. These models, which haven't been studied in much detail yet, have the potential of realizing intriguing phases such as chiral symmetry breaking without confinement, free magnetic phases, and mixed phases. $S$-duality makes these phenomena manifest in the perturbative (but not UV complete, as we discuss below) magnetic description. The weakly coupled duals can then be used to shed light on the strong dynamics of gauge theories and their possible phases.
 
Admittedly, the combination of strong dynamics and absence of supersymmetry
makes it hard to provide strong analytic evidence of the duality beyond the matching of global anomalies and the arguments from string theory. On the other hand, recent progress on lattice techniques for fermions in two-index representations suggest that these theories could be analyzed numerically.
The possibility of a nonperturbative confirmation of the duality is extremely interesting. On the field theory side, it would allow for a more detailed understanding of the strong dynamics of nonsupersymmetric theories and their phases. For string theory,
 it would provide an explicit test of $S$-duality in a nonsupersymmetric system, going beyond the current checks of the duality.

The paper is structured as follows. First, \S \ref{sec:string} explains the string theory origin of the nonsupersymmetric $S$-dualities. In \S \ref{sec:nonsusyS} we analyze concrete examples of dualities, coming from branes in flat space and orbifold singularities, and briefly discuss other possible settings. \S \ref{sec:QFT} presents a field theory analysis of a duality between a strongly coupled $SU(4)$ gauge theory and a magnetic $SO(5)$ theory. This is a simple example that illustrates many of the properties discussed before, including fermions in multiple representations and the possibility of new IR phases. Finally, \S \ref{sec:concl} contains our conclusions and future directions.

\section{String theory origin}\label{sec:string}

Let us begin by describing the string theory setup  -- brane ingredients and internal geometry -- for nonsupersymmetric $S$-dualities.  We will start with a supersymmetric gauge theory living on D3 branes and then break supersymmetry explicitly by replacing the color branes by antibranes.

Consider a four dimensional supersymmetric gauge theory with gauge group $G$, realized on the worldvolume of $N$ D3 branes. We will add orientifold planes (O3 and O7 planes in our examples below), which project $SU$ gauge groups down to $SO$ or $Sp$. The orientifolds play an important role, allowing for dualities between different gauge groups where $SO$ and $Sp$ factors are exchanged -- as is familiar from the Montonen-Olive duality.\footnote{It would also be interesting to study nonsupersymmetric self-dual theories.} Other brane ingredients may include D5 or D7 branes.  Their effects on the gauge theory depend on whether their extra dimensions wrap compact or noncompact cycles along the internal space. The later give rise to flavors in the field theory. Our examples will make use of a particular type of compact branes -- fractional branes -- as we will explain shortly.

The D3 branes can probe different 6d internal spaces of the 10d type IIB theory, whose geometric properties are encoded in the worldvolume theory on the branes. The internal space has to be much larger than the scales of interest in the gauge theory so that the coupling to gravity can be neglected; near the D3 branes the internal space can be treated effectively as noncompact. Geometric singularities, such as orbifolds $\mathbb R^6/\Gamma$, are especially interesting: 
a large number of gauge theories can be realized on D-branes near singularities, including examples very similar to the Standard Model~\cite{Aldazabal:2000sa}. As the singularity is approached, certain cycles shrink to zero size. D5 and D7 branes can wrap vanishing 2- and 4-cycles, respectively. These configurations are nonsingular in the presence of worldvolume magnetic flux and are known as fractional branes~\cite{Diaconescu:1997br}. Because of the magnetic flux, they carry D3 charge and hence they contribute to the gauge group rank.

Having explained the basic ingredients, we consider a supersymmetric gauge theory with gauge group $G$ that admits a known
$S$-dual description. The D3 branes are the same on both sides of the duality, but generically the dual
gauge group $G'$, matter content and interactions will be different. (Primes are used to distinguish between the two $S$-dual theories).  The action of $S$-duality in the type IIB string theory is described in~\cite{Polchinski:1998rr,Witten:1998xy}.  We will require that both of these gauge theories admit a Lagrangian formulation, which turns out to be a strong constraint. For instance, the action of $S$ on flavor 5- and 7-branes in general leads to strongly coupled theories for which it is hard to find a Lagrangian description. For this reason, in the present work we restrict to fractional branes. These theories can be efficiently studied using the techniques of~\cite{Franco:2005rj}.

Now we come to the key step, replacing the $N$ D3 branes by $\overline{D3}$ branes. In general this breaks all the supersymmetries. As we illustrate in \S \ref{sec:nonsusyS}, the following simple rules take into account the effect of reversing the sign of the RR charge:
\begin{itemize}
\item $N$ is replaced by $-N$ in the rank of the gauge group; for instance a rank $|N+k|$ changes to $|N-k|$.
\item Symmetrization and antisymmetrization of worldvolume fermions is exchanged, while the bosons are not modified.
\end{itemize}
Some comments are in order here. First, recall that the rank of the gauge group receives contributions both from D3 and fractional branes. A physical way of distinguishing them is by going along the Coulomb branch; this reveals the $N$ mobile D3 branes --the fractional branes cannot be moved away from the singularity because their tension would increase. Performing the replacement $N \to -N$ requires identifying these two sources. Furthermore, our theories have orientifold planes, which project some of the fields in the adjoint down to a symmetric or antisymmetric representation. When the sign of the RR charge is reversed, there is a relative minus sign in the orientifold action on bosons and fermions~\cite{Sugimoto:1999tx}, explaining the second rule. 

Switching $D3 \to \overline{D3}$ in the theory $G$ above gives a new $\t G$, and similarly $G' \to \t G'$, according to the rules we just discussed. Since both $D3$ and $\overline{D3}$ branes are self-dual, we obtain a nonsupersymmetric $S$-duality between $\t G$ and $\t G'$. Another way of viewing this duality is as follows. We start from the supersymmetric $S$-dual pair $G$ and $G'$, with $N$ D3 branes, and  add $2N$ $\overline {D3}$ branes to each side of the duality. The duality should still hold because the antibranes are selfdual. The color branes annihilate against the antibranes, leaving a duality between nonsupersymmetric gauge theories on the worldvolume of the remaining $N$ $\overline{D3}$ branes. 

An important property of the theories analyzed in this work is that the source of supersymmetry breaking is the previous exchange of fermionic representation (after the brane-antibrane annihilation takes place).
In other words, the gauge theories would be supersymmetric were it not for this modification. This is related to the fact that there are no noncompact flavor branes, which would have introduced additional sources of supersymmetry breaking. Therefore,
having antibranes amounts to a nonsupersymmetric orientifold projection that treats bosons and fermions differently. We will see this nonsupersymmetric projection at work in an $SU(N)^3$ quiver in \S \ref{sec:nonsusyS}.

We also need to specify the interactions in the nonsupersymmetric case. Since the breaking of supersymmetry comes from a nonsupersymmetric orientifold projection, the tree level potential is taken to be the same as in the supersymmetric theory, with the modification that the fermionic representations are reversed as before. The structure of the potential is important because these theories contain scalar fields inherited from the flat directions of the supersymmetric parent Lagrangian. 
The scalars are massless at tree level, but will generically be lifted by loop corrections. The condensation of some of these scalars will play a central role, providing a dual realization of chiral symmetry breaking.

We have described our mechanism as starting from a known $S$-dual pair of supersymmetric theories, and then replacing color branes by antibranes. In principle, the step of going through the supersymmetric system is not strictly required. Given that the type IIB $S$-duality is believed to be exact, one could try to construct directly nonsupersymmetric configurations. However, the action of $S$-duality is not always easy to determine (particularly near singularities), so it is helpful to start from a supersymmetric setup where the duality can be tested convincingly, and then perform the replacement $N \to -N$ together with the nonsupersymmetric orientifold projection. It would nevertheless be interesting to analyze $S$-dual theories with additional sources of supersymmetry breaking.

\subsection{Subtleties and limitations}

There are various concerns that come up when considering nonsupersymmetric dualities, which are worth emphasizing. At the level of the field theory analysis, the dynamics becomes intractable beyond weak coupling, and it is hard to provide strong analytic checks of the duality. Nevertheless, one piece of evidence for the duality will come from the matching of global anomalies for arbitrary gauge group ranks.

The main problem that we would like to point out
regards the stabilization of the scalars that we discussed before. We will study their fate at one loop, finding cases where the scalars are stabilized at the origin and others where the scalars obtain tachyonic masses. However, $S$-duality involves strong dynamics and an analytic calculation of the quantum corrections to the scalar masses becomes impossible.  In the examples of \S \ref{sec:nonsusyS}, the strongly coupled electric theory has scalars that are massive at one loop, while the magnetic description has tachyonic instabilities. In the electric theory it is in general impossible to determine what happens to the scalars in the IR due to uncontrolled corrections from strong coupling. Fortunately, this is the regime that can be described by the weakly coupled $S$-dual. 

Matching global symmetries across the duality, we find that the perturbative instability in the magnetic theory is dual to chiral symmetry breaking in the electric theory. However, given the current level of understanding of the duality, the magnetic theory cannot predict the precise pattern of symmetry breaking. The reason is that, even if the tachyon appears self-consistently at weak coupling in the regime of interest, the endpoint of the instability occurs near the UV cutoff (which could be the string scale) and is thus sensitive to the UV completion. Therefore, while we can list all of the possible symmetry breaking patterns and analyze their dynamical consequences, at this stage we do not know which one is realized. We hope that this important limitation is not fundamental, and that
a more detailed analysis of the embedding in string theory can clarify this point. In any case, the duality predicts that chiral symmetry will be broken. This turns out to have interesting consequences, particularly for theories at the borderline between confinement and conformality.

\section{Nonsupersymmetric S-duality}\label{sec:nonsusyS}

The simplest realization of this mechanism is on antibranes in flat 6d space~\cite{Uranga:1999ib,Sugimoto:2012rt}. After reviewing this example, we present dualities from orbifold singularities and discuss other possible generalizations. The dual theories obtained in this way have in general product gauge groups and fermions in multiple two-index representations.

\subsection{Nonsupersymmetric $S$-duality in $\mathbb R^6$}\label{subsec:R6}

It is instructive to first review the simplest case, studied in~\cite{Uranga:1999ib,Sugimoto:2012rt}, corresponding to the $O3-\overline{D3}$ configuration in $\mathbb R^6$. We begin with the supersymmetric gauge theory and then replace the branes by antibranes.

The theory $G$ (in our previous notation) is obtained from $2N$ D3 branes on top of an $O3^+$ orientifold plane,\footnote{After orientifolding, we have $N$ D3 branes and their orientifold images.} placed in a transverse space $\mathbb R^6$. The gauge theory is
\begin{center}
\be
\label{eq:electricN4}
\begin{tabular}{c|c|c}
&$Sp(2N)_G$&$SO(6)$\\
\hline
&&\\[-8pt]
$A_\mu$ &  $\symm$ & 1  \\
$\psi$ & $\symm$ & 4 \\
$\phi$ & $\symm$ & 6
\end{tabular}
\ee
\end{center}
The gauge group is $Sp(2N)_G$, with $Sp(2) \simeq SU(2)$, and the subscript `$G$' distinguishes it from the global flavor symmetries. In this work, the fermions are always in the Weyl representation. The relevant interactions are uniquely fixed by supersymmetry and the $SO(6)$ symmetry.\footnote{In $\mc N=1$ language, the six scalars and three of the Weyl fermions can be combined into three chiral superfields $\Phi_i$ while the remaining fermion (the gaugino) and gauge field give a vector multiplet. The interactions are then given by the superpotential $W =h \epsilon_{ijk} \Phi_i \Phi_j \Phi_k$ and the usual gauge interactions, with $h$ proportional to the gauge coupling.} The Coulomb branch is specified by the eigenvalues of the six scalars $\phi$ (modulo permutations); geometrically they describe the positions of the D3 branes along the internal $\mathbb R^6$. The $SO(6)$ symmetry is the rotational symmetry of the internal space.

The type IIB $S$-duality transforms the $O3^+$ plane into an $\widetilde{O3^-}$, which can be thought of as an $O3^-$ together with a stuck D3 brane~\cite{Witten:1998xy}. The $O3^-$ projects the gauge group to SO, so taking into account the 1/2 D3 brane gives the matter content for the theory $G'$
\begin{center}
\be
\label{eq:magneticN4}
\begin{tabular}{c|c|c}
&$SO(2N+1)_G$&$SO(6)$\\
\hline
&&\\[-8pt]
$A_\mu'$ &  $\asymm$ & 1  \\
&&\\[-8pt]
$\psi'$ & $\asymm$ & 4 \\
&&\\[-8pt]
$\phi'$ & $\asymm$ & 6
\end{tabular}
\ee
\end{center}
where the relevant interactions are again uniquely fixed by symmetries.
Therefore, string $S$-duality reproduces Montonen-Olive duality between $Sp(2N)$ and $SO(2N+1)$ $\mc N=4$ SYM.

Now we replace the D3 branes by antibranes. The theory (\ref{eq:electricN4}) becomes
\begin{center}
\be
\label{eq:electricN4nonsusy}
\begin{tabular}{c|c|c}
&$Sp(2N)_G$&$SO(6)$\\
\hline
&&\\[-8pt]
$A_\mu$ &  $\symm$ & 1  \\
$\psi$ & $\asymm$ & 4 \\
$\phi$ & $\symm$ & 6
\end{tabular}
\ee
\end{center}
The tree level interactions are inherited from the $\mc N=4$ theory.  Supersymmetry is explicitly broken because the fermion now transforms in the antisymmetric.\footnote{Here and in what follows, it is important to remember that the antisymmetric representation of Sp and the symmetric of SO are reducible, containing a singlet and an irreducible `traceless' part.}
 Similarly, changing the branes into antibranes in (\ref{eq:magneticN4}) gives
\begin{center}
\be
\label{eq:magneticN4nonsusy}
\begin{tabular}{c|c|c}
&$SO(2N-1)_G$&$SO(6)$\\
\hline
&&\\[-8pt]
$A_\mu'$ &  $\asymm$ & 1  \\
&&\\[-8pt]
$\psi'$ & $\symm$ & 4 \\
&&\\[-8pt]
$\phi'$ & $\asymm$ & 6
\end{tabular}
\ee
\end{center}
according to the previous rules. The rank $2N-1$ can be understood heuristically as the annihilation of the D3 stuck at the O-plane against one of the antibranes. Refs.~\cite{Uranga:1999ib,Sugimoto:2012rt} proposed that (\ref{eq:electricN4nonsusy}) and (\ref{eq:magneticN4nonsusy}) are dual descriptions of the same underlying theory. The long distance dynamics for the confining phase was studied in~\cite{Sugimoto:2012rt} making use of the weakly coupled magnetic dual.

The ``electric'' theory (\ref{eq:electricN4nonsusy}) is asymptotically free. A Coleman-Weinberg calculation shows that the scalar fields become massive at one loop -- they are no longer protected by supersymmetry. The low energy theory is an $Sp(2N)$ gauge theory with $N_f=4$ Weyl fermions in the antisymmetric, which contains a singlet and a `traceless' part. The theory for $N=1$ is pure $SU(2)$ Yang-Mills, which confines, and it is plausible that the confining phase extends at least to a finite window around this value.
In this case, presumably the $SO(6)$ global symmetry is broken to a subgroup $H$ by a Lorentz invariant fermionic condensate, leading to pions that parametrize the coset $SO(6)/H$. However, the theory for larger values of $N$ is at the borderline between confining and conformal, and lattice results on its phase structure are not conclusive.\footnote{We thank M. Unsal for discussions on this and related issues.}

The ``magnetic'' theory (\ref{eq:magneticN4nonsusy}) is IR free, providing a weakly coupled description of the long distance dynamics. The scalars acquire a tachyonic mass at one loop, condense and break the $SO(6)$ symmetry. Unfortunately, the specific pattern of symmetry breaking is not known from first principles because it depends on details of the UV completion of the theory. The simplest pattern $SO(6) \to SU(4)$ is consistent with a confining theory. But other ways of breaking the global symmetry are also possible, and they could lead to nonconfining phases. We will see examples of this in \S \ref{sec:QFT}. In any case, the duality predicts that the $SO(6)$ symmetry will be broken, a result that would be interesting to check with the current lattice simulations.

\subsection{Nonsupersymmetric $S$-duality in $\mathbb R^6/\mathbb Z_3$}\label{subsec:R6Z3}

In the previous section we illustrated the simplest version of the nonsupersymmetric $S$-duality. The approach described in \S \ref{sec:string} applies to rather general gauge theories, with the condition that they can be realized on the worldvolume of D3 branes and that the $S$-dual of the supersymmetric case is known. A generalization is to place the D-branes near singularities, instead of having them in $\mathbb R^6$ as in \S \ref{subsec:R6}. Singular geometries provide a large class of field theories with different degrees of supersymmetry and matter content. Let us focus on theories with $\mc N=1$ supersymmetry, which arise on the worldvolume of D3 branes near Calabi-Yau singularities (for a review see~\cite{Malyshev:2007zz}). Refs.~\cite{GarciaEtxebarria:2012qx,Garcia-Etxebarria:2013tba} recently analyzed $S$-duality in some of these string theory solutions and proposed dual pairs of $\mc N=1$ gauge theories, with strong evidence both from field theory and string theory. We now study the nonsupersymmetric duality obtained by replacing the D3-branes with $\overline{D3}$-branes.

The simplest theory in this class corresponds to the $\mathbb R^6/\mathbb Z_3$ orbifold, with $\mathbb Z_3$ action $z_i \to e^{2\pi i/3} z_i$ on the complex coordinates of $\mathbb C^3 = \mathbb R^6$. The orbifold breaks $\mc N=4$ to $ \mc N=1$ and $SO(6)$ to $SU(3) \times U(1)$. Placing $N$ D3 branes near this singularity gives a $SU(N)^3$ quiver gauge theory, an orbifold projection of the $\mc N=4$ $SU(3N)$ theory. In order to have a nontrivial $S$-duality, we analyze the gauge theories with orientifold planes constructed in~\cite{Franco:2010jv} using the dimer techniques of~\cite{Franco:2007ii}. 

The $\mc N=1$ ``electric'' theory comes from $2N$ D3 branes in the presence of a fractional $O7^+$ plane and 4 $\overline{D7}$ branes, and has matter content (in $\mc N=1$ notation)
\begin{center}
\be
\label{eq:electricZ3susy}
\begin{tabular}{c|cc|cc}
&$Sp(2N+4)_G$&$SU(2N)_G$&$SU(3)$&$U(1)_R$\\
\hline
&&&&\\[-8pt]
$Q$ & $\Box$& $\overline \Box$ & $\Box$&$\frac{2}{3}- \frac{1}{N}$  \\
&&&&\\[-8pt]
$S$ & $1$& $\symm$ & $\Box$&$\frac{2}{3}+ \frac{2}{N}$  \\
\end{tabular}
\ee
\end{center}
In this table, the gauge group is $G=Sp(2N+4)_G \times SU(2N)_G$, $Q$ and $S$ are chiral superfields, $U(1)_R$ is an R-symmetry, and the vector multiplets $(A_\mu,\lambda_A)$ and $(V_\mu, \lambda_V)$ of $Sp$ and $SU$ are not shown. There is an $SU(3)$ invariant superpotential $W=h \,Q S Q$.

The $S$-dual magnetic theory corresponds to $2N$ D3 branes, a fractional $O7^-$ plane, 4 D7 branes, and a D3 brane stuck at the orientifold,
\begin{center}
\be
\label{eq:magneticZ3susy}
\begin{tabular}{c|cc|cc}
&$SO(2N-1)_G$&$SU(2N+3)_G$&$SU(3)$&$U(1)_R$\\
\hline
&&&&\\[-8pt]
$Q'$ & $\Box$& $\overline \Box$ & $\Box$&$\frac{2}{3}+ \frac{2}{2N+3}$  \\
&&&&\\[-8pt]
$A$ & $1$& $\asymm$ & $\Box$&$\frac{2}{3}- \frac{4}{2N+3}$  \\
\end{tabular}
\ee
\end{center}
and a superpotential $W =h' \,Q' A Q'$. The vector multiplets $(A_\mu',\lambda_A')$ and $(V_\mu', \lambda_V')$ of the gauge groups are not shown.  
The gauge group factors have one-loop beta functions with opposite signs, both in the electric and magnetic theory. Since neither theory is asymptotically free, the field theory duality is necessarily a low energy equivalence.

Starting from this supersymmetric duality, let us change the $2N$ D3 branes into antibranes.  Applying the rules in \S \ref{sec:string} to the electric theory gives\footnote{To summarize our notation, in the rest of the work $A_\mu$ and $V_\mu$ denote gauge fields, $\lambda$ and $\psi$ are Weyl fermions, and other letters denote complex scalars.}
\begin{center}
\be
\label{eq:electricZ3nonsusy}
\begin{tabular}{c|cc|cc}
&$Sp(2N-4)_G$&$SU(2N)_G$&$SU(3)$&$U(1)$\\
\hline
&&&&\\[-8pt]
$A_\mu$ & $\symm$& 1 & 1& 0  \\
&&&&\\[-8pt]
$\lambda_A$ & $\asymm$& 1 & 1& 1  \\
&&&&\\[-8pt]
$V_\mu$ & 1& adj & 1& 0  \\
&&&&\\[-8pt]
$\lambda_V$ & 1 & adj & 1& 1  \\
&&&&\\[-8pt]
$Q$ & $\Box$& $\overline \Box$ & $\Box$&$\frac{2}{3}+ \frac{1}{N}$  \\
&&&&\\[-8pt]
$\psi_Q$ & $\Box$& $\overline \Box$ & $\Box$&$-\frac{1}{3}+ \frac{1}{N}$  \\
&&&&\\[-8pt]
$S$ & $1$& $\symm$ & $\Box$&$\frac{2}{3}- \frac{2}{N}$  \\
&&&&\\[-8pt]
$\psi_S$ & $1$& $\asymm$ & $\Box$&$-\frac{1}{3}- \frac{2}{N}$ 
\end{tabular}
\ee
\end{center}
The rank of the $Sp$ factor is a consequence of the annihilation of the 4 units of D3 charge carried by the D7-branes against the same number of $\overline{D3}$. As anticipated in \S \ref{sec:string}, we can interpret (\ref{eq:electricZ3nonsusy}) as a nonsupersymmetric orientifold projection of the $SU(N)^3$ quiver, together with the replacement $N \to - N$. Note that, as in the supersymmetric case, cancellation of the gauge anomalies introduced by the orientifold requires fractional 7-branes~\cite{Franco:2010jv}.

The tree-level terms are the same as those of the supersymmetric theory, modulo the change in the fermion representations (the orientifold projection does not modify the tree-level coefficients). In particular, mass terms vanish at this order. At the renormalizable level we then have gauge, Yukawa and quartic interactions, with tree level coefficients related by the supersymmetric boundary condition. Quantum-mechanically, all the terms allowed by symmetries are expected to be generated. 

The exchange of the symmetric/antisymmetric representations for fermions when $N \to -N$ can also be understood from field theory considerations.  For instance, $\psi_S$ has to transform in the antisymmetric of the $SU$ factor in order to avoid a gauge anomaly. Switching the representation of the fermions also ensures that the 't Hooft anomaly matching conditions are still upheld between the two pairs of duals.

The magnetic description follows from replacing the $N$ D3-branes by antibranes in (\ref{eq:magneticZ3susy}):
\begin{center}
\be
\label{eq:magneticZ3nonsusy}
\begin{tabular}{c|cc|cc}
&$SO(2N+1)_G$&$SU(2N-3)_G$&$SU(3)$&$U(1)$\\
\hline
&&&&\\[-8pt]
$A_\mu'$ & $\asymm$& 1 & 1& 0  \\
&&&&\\[-8pt]
$\lambda_A'$ & $\symm$& 1 & 1& 1  \\
&&&&\\[-8pt]
$V_\mu'$ & 1& adj & 1& 0  \\
&&&&\\[-8pt]
$\lambda_V'$ & 1 & adj & 1& 1  \\
&&&&\\[-8pt]
$Q'$ & $\Box$& $\overline \Box$ & $\Box$&$\frac{2}{3}- \frac{2}{2N-3}$  \\
&&&&\\[-8pt]
$\psi_Q'$ & $\Box$& $\overline \Box$ & $\Box$&$-\frac{1}{3}- \frac{2}{2N-3}$  \\
&&&&\\[-8pt]
$A$ & $1$& $\asymm$ & $\Box$&$\frac{2}{3}+ \frac{4}{2N-3}$ \\
&&&&\\[-8pt]
$\psi_A$ & $1$& $\symm$ & $\Box$&$-\frac{1}{3}+ \frac{4}{2N-3}$ 
\end{tabular}
\ee
\end{center}
and, as before, the tree level potential is obtained by requiring a supersymmetric theory if the representations of the fermion were changed; radiative corrections generate all the terms allowed by symmetries. The rank of the $SO$ factor combines the $N$ $\overline{D3}$s and the 4 fractional $D7$ branes, which annihilate against the stuck D3 charge; there is a similar cancellation in the $SU$ factor, except that the fractional 7-branes do not contribute.

We propose that (\ref{eq:electricZ3nonsusy}) and (\ref{eq:magneticZ3nonsusy}) are $S$-dual theories. 
A (weak) check of the duality is that 't Hooft's anomaly matching is satisfied for all $N$. In fact, the matching of global anomalies is inherited from the supersymmetric duality. 

As presented, the duality contains scalar fields; these are massless at tree level but, in contrast with the supersymmetric case, they obtain masses from quantum corrections. A one-loop string theory calculation generically gives $m^2 \sim g_s/\alpha'$, where $g_s$ is the string coupling and $\alpha'^{1/2}$ is the string scale~\cite{Uranga:1999ib}. In field theory terms, $g_s$ determines the strength of the gauge, Yukawa and quartic couplings (e.g. $g_\text{YM}^2 \sim g_s$), and $\alpha'^{-1/2}$ is the natural UV cutoff. The electric theory contains more scalars than fermions, so we expect the scalars to acquire positive one loop masses. The situation is the opposite in the magnetic theory, where the scalars would obtain tachyonic masses and break the gauge and global symmetries. 

We now make some prelimiary remarks on the dynamics of the electric theory. Let us assume that the one-loop result continues to hold at strong coupling, so that the scalar fields are massive. Below this mass scale we are left with a gauge theory that only contains fermions and, according to the one loop beta functions of the gauge couplings, both gauge group factors are asymptotically free. Therefore, this theory is UV complete by itself; this should be contrasted with the supersymmetric case, where the gauge group factors had beta functions of opposite sign. In order to develop some intuition on the IR dynamics, we can compute the two loop beta functions. (See e.g.~\cite{Caswell:1974gg}). For the $Sp$ subgroup, the one and two loop factors have opposite sign, suggesting an IR fixed point. On the other hand, the two loop factor for the $SU$ subgroup changes sign around $N \approx 3$. Based on this approximation, $N=2$ would be the boundary of the conformal window. Of course, these calculations are not
under analytic control, and the long distance dynamics may be different. The case $N=2$
 will be studied in more detail in \S \ref{sec:QFT}.

We end this general analysis by noting the following interesting phenomenon. Integrating out the electric scalars at one loop yields three different anomaly free $U(1)$ symmetries, from the five different fermions ($\lambda_A$ is in a reducible representation). If these symmetries were exact, they would lead to
additional 't Hooft anomalies that are not matched across the duality. Therefore, the duality predicts the existence of dangerously irrelevant operators which will break the extra $U(1)$ symmetries in the electric theory. These operators are known to exist in supersymmetric gauge theories, but in a nonsupersymmetric setup it is very hard to establish their existence. Here they are required by $S$-duality and the matching of global symmetries.

\subsection{Generalizations}\label{subsec:generalizations}

There are various generalizations that can be considered. A direct extension of \S \ref{subsec:R6Z3} is to orbifolds $\mathbb R^6/\mathbb Z_{2k+1}$ with $k>1$. These are quiver gauge theories with $k+1$ nodes.  Refs.~\cite{Bianchi:2013gka,Garcia-Etxebarria:2013tba} conjectured a supersymmetric $S$-duality between
\be
\prod_{a=1}^k SU(N_a) \times Sp(N_{k+1})\;,\;N_a \equiv 2N + 4 \left\lfloor \frac{a}{2} \right\rfloor
\ee
and
\be
\prod_{a=1}^k SU(N_a') \times SO(N_{k+1}')\;,\;N_a'\equiv2N+2k+1 - 4 \left\lfloor\frac{a}{2}\right\rfloor
\ee
with a matter content that is a generalization of the one for $k=1$ discussed above. Replacing D3-branes with antibranes, $N \to -N$, gives an infinite family of nonsupersymmetric dual theories.

Another possibility is to consider singularities that are not orbifolds. Complex cones over del Pezzo surfaces $dP_n$ are well-known examples; the singular limit of $dP_0$ is the orbifold $\mathbb C^3/\mathbb Z_3$ analyzed before, but higher del Pezzos give rise to non-orbifold singularities. D3 branes probing these 6d spaces (and appropriate fractional D5, D7 and orientifold planes) give rise to $\mc N=1$ gauge theories. Supersymmetric $S$-dualities on some of these spaces have been discussed in~\cite{GarciaEtxebarria:2012qx}.
In particular, we propose that antibranes probing orientifolds of $dP_1$ give rise to an $S$-duality between nonsupersymmetric theories
\be
SU(2N-5) \times SU(2N-1) \,\leftrightarrow\,SU(2N+1) \times SU(2N-3)
\ee
which we plan to study in more detail in the future.

It would also be interesting to study the nonsupersymmetric dualities inherited from $\mc N=2$ theories, which could have qualitatively different properties than the $\mc N=1$ examples analyzed here. A class of theories that may be tractable arise from 3-branes probing orbifolds of the form
$\mathbb R^4/\mathbb Z_n \times \mathbb R^2$. 

Finally, a qualitatively different class of theories may arise from D3-branes probing nonsupersymmetric singularities. The simplest possibility would be a nonsupersymmetric orbifold of $\mathbb R^6$, suggested also in~\cite{GarciaEtxebarria:2012qx}. Understanding orientifolds and
the action of the string $S$-duality in these systems may lead to new nonsupersymmetric dualities.

\section{S-duality for a simple Lie group}\label{sec:QFT}
 
This last part of the paper is devoted to a more detailed field theory analysis of a specific nonsupersymmetric duality. As we saw in the previous section, the theories that appear naturally from antibranes and orientifolds at singularities generically have product gauge groups. This makes an explicit field theory analysis somewhat involved; in order to understand the dynamics in a simpler setup, we would like to have dualities involving simple gauge groups. 

With this motivation, let us analyze
 the duality between (\ref{eq:electricZ3nonsusy}) and (\ref{eq:magneticZ3nonsusy}) for $N=2$, namely two $\overline{D3}$ and their orientifold images. For this value of $N$ the duality becomes particularly simple and predicts the low energy equivalence between an electric $SU(4)$ and a magnetic $SO(5)$ gauge theories. These theories feature fermions in multiple two-index representations, a property that can lead to a rich strongly coupled dynamics. This will be reflected in perturbative properties of the magnetic description.

\subsection{Electric Theory}\label{subsec:QFTelectric}

Setting $N=2$ in (\ref{eq:electricZ3nonsusy}) gives the electric theory
\begin{center}
\be
\label{eq:Electric}
\begin{tabular}{c|c|ccc}
&$SU(4)_G$&$SU(3)$&$U(1)$&$\mathbb{Z}_4$\\
\hline
&&&\\[-8pt]
$V_\mu$ &  adj &$1$&$0$&0  \\
&&&\\[-8pt]
$\lambda_V$ & adj &$1$ &$1$&$w_4$\\
&&&\\[-8pt]
$S$ & $\symm$&$\fund$&$-\frac{1}{3}$&$w_4$\\
&&&\\[-8pt]
$\psi_S$ & $\asymm$&$\fund$&$-\frac{4}{3}$&$0$
\end{tabular}
\ee
\end{center}
where $w_4 = \exp{2 \pi i/4}$.  For ease of future discussion, we included a $\mathbb{Z}_4$ discrete group which is a combination of the center of $SU(3)$ and the $U(1)$.  The $Sp$ node of (\ref{eq:electricZ3nonsusy}) and all the fields charged under it disappear for $N=2$.  Global symmetries forbid fermion masses, so gauge interactions give the only renormalizable terms in the Lagrangian. The global symmetries will also be important for matching gauge invariants with the dual description. We have included the scalar $S$ in (\ref{eq:Electric}) because it is massless at tree level, but it receives a positive one-loop mass and is expected to be lifted.

This theory is asymptotically free, becoming strongly coupled at low energies. The IR phase structure of nonsupersymmetric theories is not fully understood. In this case, a simple guess is that since the first two coefficients in the perturbative gauge coupling beta function have the same sign, the theory may confine. This is, however, an uncontrolled approximation, so the IR phase could be different.
Based on intuition from QCD, a fermion bilinear may also condense and break the global symmetry. In fact, the magnetic dual will predict that chiral symmetry breaking occurs.

The appearance of fermions $\lambda_V$ and $\psi_S$ in different two-index representations
suggests a rich nonperturbative dynamics. In particular, the condensation of the fermion bilinears
$\lambda_V \lambda_V$ and $\psi_S \psi_S$ can lead to quite different patterns of chiral symmetry breaking and low energy theories of pions.  While a condensate for $\lambda_V \lambda_V$ breaks only the $U(1)$,  condensation of $\psi_S \psi_S$ would break both $U(1)$ and $SU(3)$, but not the $\mathbb{Z}_4$.  Depending on the nonperturbative dynamics, the $SU(3)$ symmetry can be broken in different ways, the simplest possibility being $SU(3) \to SO(3)$.  Multiple fermionic representations also offer the possibility of dissociating  chiral symmetry breaking and confinement, as the magnetic description below suggests.

It is interesting to note that there has been recent progress on lattice results for gauge theories with matter content similar to the one that has appeared here. See~\cite{Armoni:2008nq,DelDebbio:2009fd} for some examples and additional references. Given these developments, it seems reasonable that theories like (\ref{eq:Electric}) could be studied on the lattice. We should stress that (as explained before) the scalar $S$ is expected to be lifted by quantum effects, so it need not be included in a numerical calculation.
It would then be possible to test the predictions of nonsupersymmetric $S$-duality, to which we turn next.

\subsection{Magnetic Theory}\label{subse:QFTmagnetic}

The dual magnetic theory from (\ref{eq:magneticZ3nonsusy}) with $N=2$ simplifies to
\begin{center}
\be
\label{eq:Magnetic}
\begin{tabular}{c|c|ccc}
&$SO(5)_G$&$SU(3)$&$U(1)$&$\mathbb{Z}_4$\\
\hline
&&&\\[-8pt]
$A_\mu'$  & $\asymm$ & $1$&$0$&0 \\
&&&\\[-8pt]
$\lambda_A'$ & $\symm$ & $1$&$1$& $w_4$\\
&&&\\[-8pt]
$Q'$ & $\fund$ &$\fund$ &$-\frac{4}{3}$ & 0\\
&&&\\[-8pt]
$\psi_Q'$ & $\fund$ &$\fund$ &$-\frac{7}{3}$ & $w_4^{-1}$\\
&&&\\[-8pt]
$\psi_A$ & 1 & $\fund$&$\frac{11}{3}$ & $w_4$ \\
\end{tabular}
\ee
\end{center}
\bea
\nonumber
\eea
The scalar field $Q'$ is massless at tree level, and its one-loop mass will be studied shortly. The tree level interactions inherited from the supersymmetric model include the usual gauge interactions, Yukawa terms and quartics from D-terms; the superpotential interactions vanish for $N=2$.

The $SO(5)_G$ gauge group is asymptotically free. However, a cutoff $\Lambda$ (which could be the string scale) appears in the quantum corrections to the mass of the scalar. In the supersymmetric case, the mass corrections cancel exactly between the bosonic and fermionic contribution. In contrast, the nonsupersymmetric theory has more fermions than scalars,
so  $Q'$ obtains a negative mass squared at one-loop, and thus induces the spontaneous symmetry breaking of $SO(5)_G \times SU(3) \times U(1)$. The tachyonic mass is of order $m^2 \sim - g_s \Lambda^2$, so if the stabilization occurs due to interactions set by $g_s$, the expectation value of $Q'\sim \Lambda$. In this case the specific pattern of symmetry breaking is sensitive to the UV completion. We will discuss various possibilities shortly, and find that they also lead to quite different IR phases.

We thus obtain our first prediction from the magnetic dual: the global $SU(3) \times U(1)$ symmetry has to be broken, while the discrete $\mathbb{Z}_4$ survives. This symmetry breaking pattern suggests that the fermion bilinear $\psi_S \psi_S$ in the electric theory condenses. In fact, comparing the quantum numbers on both sides, this fermion bilinear appears to have a simple interpretation in terms of $Q' Q'$ in the magnetic theory. The expectation value of $Q'$ leads to massive spin one fields from the Higgsing of $SO(5)_G$. These would play the role of rho mesons of the electric theory.\footnote{However, given what we currently know about the duality, it is not clear what the mass scale for the rho mesons is in the electric theory.}  A similar behavior is observed in QCD, where the rho mesons can be interpreted as coming from an emergent gauge group, and the fermion condensate maps to a product of Higgs fields which break the emergent symmetry; see e.g.~\cite{Komargodski:2010mc}  for a recent discussion and references. 

\subsection{Symmetry Breaking Patterns}\label{subsec:patterns}

Let us consider the pattern of symmetry breaking in the magnetic theory.  We first note that from~(\ref{eq:Electric}), $\psi_S$ is uncharged under a discrete $\mathbb{Z}_4$ symmetry while $\lambda_V$ is charged.  Thus $\mathbb Z_4$ provides a useful handle for discerning which fermion bilinears condense. One important consequence of this symmetry is that if it is unbroken, 't Hooft anomaly matching of the (gravity)$^2 \mathbb{Z}_4$ anomaly requires the existence of at least one massless fermion~\cite{Csaki:1997aw}.

$Q'$ has a negative mass that is $m^2 \sim -g_s \Lambda^2$ and develops a vev around the cutoff $Q'  \sim \Lambda$.  Thus both renormalizable and nonrenormalizable terms become important for the stabilization of this direction. Nevertheless, it is useful to develop some intuition by restricting to a renormalizable potential.  $Q'$ has two indices, a gauge index and a flavor index.  In what follows, we consider $Q'$ to be a $5 \times 3$ matrix.  The renormalizable potential is 
\bea
V &=&  - m^2 \text{Tr} \left [ Q' Q'^\dagger \right ] + \lambda_1 \text{Tr} \left [ Q' Q'^\dagger \right ]^2\nonumber \\
  &+& \lambda_2 \text{Tr} \left [ Q' Q'^\dagger Q' Q'^\dagger \right] + \lambda_3 \text{Tr} \left [ Q'^T Q' Q'^\dagger Q'^* \right ]
\eea
Using $SU(3)$ symmetry rotations, the $3 \times 3$ matrix $Q'^\dagger Q'$ can be diagonalized.  $SO(5)_G$ symmetry rotations are not enough to diagonalize the $5 \times 5$ matrix $Q' Q'^\dagger$.  Thus we have expectation values of the form
\bea
Q' = \left( \begin{array}{ccc}
i x & 0 & 0 \\
0 & i y & 0\\
a & 0 & 0\\
0 & b & 0\\
0 & 0 & c \end{array} \right)
\eea
for real $a,\,b,\,c,\,x,\,y$.  Because the gauge group is larger than the flavor symmetry, it is possible that the hidden local symmetry will not be completely Higgsed.  

There are many symmetry breaking patterns available depending on the relative signs and sizes of the three couplings.  As an example, consider the supersymmetric boundary conditions, $\lambda_1=0$ and $\lambda_2=-\lambda_3 \sim g^2$.  These SUSY boundary conditions have flat directions that are lifted by radiative corrections.  Suppose that the radiative corrections and/or higher dimensional operators lift the flat directions by increasing all of the couplings slightly.  These conditions uniquely specify the symmetry breaking pattern to be $SO(5)_G \times SU(3) \times U(1) \,\to\,SO(2)_G \times SO(3) \times \mathbb{Z}_4$ corresponding to the expectation values $x=y=0$ and $a=b=c\ne0$.  The matter content is
\begin{center}
\be
\begin{tabular}{c|c|cc}
&$SO(2)_G$&$SO(3)$&$\mathbb Z_4$\\
\hline
&&\\[-8pt]
$A_\mu'$  & $\asymm$ & $1$ & $0$\\
&&&\\[-8pt]
$\lambda_A'$ & $\symm$ & $1$ & $w_4$\\
&&&\\[-8pt]
6 GB && $SU(3) \times U(1)/SO(3)$&\\
\end{tabular}
\ee
\end{center}
The allowed interaction $Q' \psi_Q' \psi_A$ lifts $\psi_A$ and three of the $\psi_Q'$ components while the coupling $Q'^\dagger \lambda_A' \psi_Q'$ lifts the remaining components of $\psi_Q'$.
This symmetry breaking pattern is natural from the electric theory perspective, as it corresponds to the most symmetric expectation value for the fermion bilinear $\psi_S \psi_S$ (assuming that it condenses).  Strikingly,
if this turns out to be the symmetry breaking pattern that is realized, the magnetic theory predicts an emergent gauge symmetry with massless fermions. The theory is then in a free magnetic phase with chiral symmetry breaking. Free magnetic phases are known in supersymmetric theories~\cite{Intriligator:1995au}, but we are not aware of models where this phase occurs together with the breaking of chiral symmetry in a theory with a simple gauge group.

Stability of the potential requires that $\lambda_1 + \lambda_2 + \lambda_3 > 0$.  Another interesting scenario is the maximally stable one where $\lambda_{1,2,3} > 0$.  The symmetry breaking pattern for these values of $\lambda$ is $SO(5)_G \times SU(3) \times U(1)\,\to\,SO(2) \times \mathbb Z_4$ with $x=y=a=b\ne0$ and $c\ne0$.  The matter content is
\begin{center}
\be
\begin{tabular}{c|cc}
&$SO(2)$&$\mathbb Z_4$\\
\hline
&\\[-8pt]
$\lambda_A'$ & $\symm$ & $w_4$ \\
&\\[-8pt]
8 GB &$SU(3) \times U(1)/SO(2)$&\\
\end{tabular}
\ee
\end{center}
We have the Goldstone bosons required by the symmetry breaking pattern, but also three massless fermions transforming in a triplet of the surviving global symmetry, as required by discrete anomaly matching. This symmetry breaking pattern leads to confinement with chiral symmetry breaking.

There are also more exotic symmetry breaking patterns available.  For instance, for $\lambda_1 > 0$ and $\lambda_{2,3} <0$, the symmetry breaking pattern is $SO(5)_G \times SU(3) \times U(1)\,\to\,SO(4)_G \times SU(2) \times U(1)$, with $x=y=a=b=0$ and $c\ne0$.  The matter content of this theory is
\begin{center}
\be
\begin{tabular}{c|c|cc}
&$SO(4)_G$&$SU(2)$&$U(1)$\\
\hline
&&\\[-8pt]
$A_\mu'$  & $\asymm$ & $1$ & $0$\\
&&\\[-8pt]
$\lambda_A'$ & $\symm$ & $1$ & $1$\\
&&\\[-8pt]
$\psi_Q'$ & $\fund$ & $\fund$ & $-3$ \\
&&\\[-8pt]
$\psi_A$ & 1 & 1 & $5$ \\
&&\\[-8pt]
5 GB && $SU(3)/SU(2)$\\
\end{tabular}
\ee
\end{center}
The $SO(4)_G$ is asymptotically free so the IR of this theory becomes strongly coupled.  Even this more exotic symmetry breaking pattern is interesting.  The one and two loop beta functions for the gauge coupling now have opposite sign, so it is plausible that the theory flows to an interacting fixed point and the gauge neutral fields decouple. This is a nonsupersymmetric mixed phase, which also has an analog in supersymmetric theories~\cite{Craig:2011tx}.
The presence of an emergent gauge group becoming strongly coupled suggests that multiple scales would be involved in the confinement of the electric theory.  This sort of behavior is of interest to technicolor model building as a way to explain the hierarchies in the Yukawa couplings.

To summarize, we find that, regardless of the symmetry breaking pattern, interesting physics is predicted by the duality.  If the symmetry breaking pattern coincides with the simplest intuition from the electric theory, there is a free magnetic phase with chiral symmetry breaking.  Other symmetry breaking patterns feature massless fermions and mixed phases.  It would be very interesting to simulate the theory on the lattice to find which of the symmetry breaking patterns is the correct one.

\subsection{Mass deformations}

Mass deformations can be treated as spurions and mapped between the electric and magnetic theory.  Consider adding the spurion mass terms, 
\begin{center}
\be
\begin{tabular}{c|ccc}
&$SU(3)$&$U(1)$&$\mathbb{Z}_4$\\
\hline
&&&\\[-8pt]
$m$ &$1$&$-2$&$w_4^2$ \\
&&&\\[-8pt]
$\tilde m$ &$\symm$&$\frac{8}{3}$&$0$
\end{tabular}
\ee
\end{center}
The electric theory has the potential
\bea
\mathcal{L}_{\text{elec.}} \supset m \lambda_V \lambda_V + \tilde m \psi_S \psi_S
\eea
As long as $m, \tilde m \ll \Lambda_{\text{dyn}}$, the duality should still hold and we can use a spurion analysis to determine the zeroth order effects on the magnetic theory.

The effects on the magnetic theory at the renormalizable level are
\bea
\label{Eq: mass terms}
\mathcal{L}_{\text{mag.}} \supset  m \lambda'_A \lambda'_A + \tilde m \Lambda \,{\rm Tr}( Q' Q' ) + h.c.
\eea
where $\mathcal{O}(1)$ constants have been left out.  We see explicitly that as masses become larger than the dynamical scale, there will be phase transitions invalidating the duality.  The one-loop negative mass of $Q'$ comes from $\lambda'_A$ and integrating it out gives $Q'$ a 1-loop positive mass instead.  The second term in (\ref{Eq: mass terms}) directly affects the symmetry breaking pattern for large $\tilde m$.

To this order, we have the mapping $\lambda_V \lambda_V \sim \lambda'_A \lambda'_A$ and $\psi_S \psi_S \sim \Lambda Q' Q'$.  The first identification is reminiscent of the mapping $W_\alpha W^\alpha \sim W'_\alpha W'^\alpha$ for supersymmetric theories.  The second is the mapping of a fermion bilinear to a scalar squared which, as we discussed before, may also explain some of the properties of QCD.

The first term in (\ref{Eq: mass terms}) gives a mass to the fermions $\lambda'_A$ while the second gives a mass to the pions.  The mass given to the pions depends on the symmetry breaking pattern.  As an example, consider the symmetry breaking pattern $SU(3) \,\to\, SU(2)$ and its 5 pions.  Giving all the electric quarks equal masses means that the symmetric breaking pattern becomes $SO(3) \,\to\, SO(2)$; 2 of the 5 pions will remain massless despite their constituent quarks obtaining non-zero mass.  This somewhat unintuitive scaling of the pion masses persists for all symmetry breaking patterns except for $SU(3) \times U(1) \,\to\, SO(3) \times \mathbb{Z}_4$.  


\section{Conclusions and future directions}\label{sec:concl}

In this work we have constructed nonsupersymmetric $S$-dualities for four-dimensional gauge theories by replacing D3-branes with antibranes in known supersymmetric $S$-dual pairs. Large classes of chiral and vector-like theories can be obtained in this way, and the duality was illustrated on 3-branes probing geometric singularities.
This method is relevant for understanding the dynamics of QCD-like theories as well as more intriguing phases such as a free magnetic phase, or mixed phases where an interacting fixed point coexists with a decoupled neutral sector. String theory offers various tools for understanding the dynamics of nonsupersymmetric gauge theories, and there are many directions for further developments.

The main limitation of the duality so far is that the magnetic description is not complete by itself -- the pattern of symmetry breaking depends on details of the UV completion within string theory. It would be important to understand these effects. It is also necessary to construct a precise dictionary between dual variables, generalizing the mapping between fermion condensates and tachyonic scalars that we found.

This method opens up the possibility of studying dualities of nonsupersymmetric theories with different matter content and interactions. Here we focused on cases where the parent supersymmetric theories have $\mc N=4$~\cite{Uranga:1999ib,Sugimoto:2012rt} or $\mc N=1$.  New phenomena may arise for $\mc N=2$ theories. The presence of antibranes leads to a nonsupersymmetric orientifold projection, and one could consider generalizing the dimer techniques for orientifolding~\cite{Franco:2010jv,Franco:2007ii} to nonsupersymmetric theories. Other directions include non-orbifold singularities (we briefly discussed a possible duality for $dP_1$) and type IIA brane systems. It would also be interesting to consider systems where supersymmetry is broken by the internal geometry and understand the action of $S$-duality.

The field theories discussed in this work also feature fermions in multiple two-index representations. These systems have not been much studied yet. As we found in \S \ref{sec:QFT} using the magnetic dual, we expect that the multiple fermionic representations lead to new phases at long distance. It would be very interesting to understand these phases in more detail.

Finally, it would be important to develop tests of the nonsupersymmetric $S$-duality, analytically or numerically. In particular, given recent developments in theories with two-index representations, the prospects of a lattice study appear encouraging. This would have important consequences for our understanding of the dynamics of gauge theories and $S$-duality in string theory.


\acknowledgements
We would like to thank
S.~Franco,
B.~Heidenreich,
K.~Intriligator,
S.~Kachru,
E.~Neil,
M.~Unsal,
A.~Uranga, and
T.~Wrase
for interesting discussions and comments on the manuscript.
A.H.~is supported by the Department of Energy under contract DE-SC0009988.
G.T.~is supported in part by the National Science Foundation under grant no.~PHY-0756174.

\end{document}